\documentclass[preprint,showpacs,preprintnumbers,amsmath,amssymb]{revtex4-1}
\usepackage{amssymb}
\usepackage{amsmath}
\usepackage{graphicx}
\usepackage{epsfig}
\usepackage{subfigure}
\usepackage{amsfonts}
\usepackage{CJK}

\begin{document}

\title{An efficient protocol of quantum walk in circuit QED}
\author{Jia-Qi Zhou}
\author{Qi-Ping Su}
\email{sqp@hznu.edu.cn}
\author{Chui-Ping Yang}
\address{Department of Physics, Hangzhou Normal University, Hangzhou, Zhejiang 311121, China}

\date{\today}

\begin{abstract}
Implementation of discrete-time quantum walk (DTQW) with
superconducting qubits is difficult since on-chip superconducting
qubits cannot hop between lattice sites.  We propose an efficient
protocol for the implementation of DTQW in circuit quantum
electrodynamics (QED), in which only $N+1$ qutrits and $N$ assistant
cavities are needed for an $N$-step DTQW. The operation of each DTQW
step is very quick because only resonant processes are adopted. The
numerical simulations show that high-similarity DTQW with the number
of step up to $20$ is feasible with present-day circuit QED
technique. This protocol can help to study properties and
applications of large-step DTQW in experiments, which is important
for the development of quantum computation and quantum simulation in
circuit QED.
\end{abstract}



\maketitle

\section{INTRODUCTION}

Circuit quantum electrodynamics (QED), composed of superconducting
qubits and microwave resonators or cavities, has attracted
substantial attention because of its controllability, integrability,
ready fabrication and potential scalability
\cite{s001,s002,s003,s01,s02,s03,s04} in quantum information and
quantum computation. The strong coupling and ultrastrong coupling of
a qubit with a microwave cavity in experiments have been reported
\cite{s09,s00}. The level spacings of superconducting qubits can be
rapidly adjusted ($1\sim3$ ns) \cite{s05,s06,s07,s08}, and their
coherence time is improved rapidly
\cite{s081,s082,s083,s084,s085,s086}. The circuit QED is considered
as one of the most feasible candidates for quantum computation and
quantum simulation \cite{s03,s04}.

Quantum walk is extension of the classical random walk, which has
wide applications in quantum algorithms \cite{q011,q02,q11}, quantum
simulation \cite{q03,q04,q041,q09}, universal quantum computation
\cite{q042,q043,q05}, and so on \cite{q07,q12}. In the standard
DTQW, there is a walker moving with respect to the state of a coin.
The evolutions of the walker and the coin are characterized by a
unitary operator $U=W\cdot C$. In each step of a 1-dimensional (1D)
DTQW, at first, the coin with states $|0\rangle_c$ and $|1\rangle_c$
is tossed by the operator
\begin{equation}
C=\cos{\theta}|0\rangle_c\langle0|+\sin{\theta}|0\rangle_c\langle1|+\sin{\theta}|1\rangle_c\langle0|-\cos{\theta}|1\rangle_c\langle1|,
 \label{coin}
\end{equation}
with $\theta\in(0,\pi/2)$, then the walker is shifted by
\begin{equation}
W=\sum_j|j+1\rangle\langle
j|\otimes|0\rangle_c\langle0|+|j-1\rangle\langle
j|\otimes|1\rangle_c\langle1|, \label{walker}
\end{equation}
where the integer $j$ represent sites of the walker in the 1D line.
The implementations of DTQW have been achieved in several quantum
systems, such as linear optics \cite{q08,q03,q11,q101,q105,q09}, ion
traps \cite{q15,q16}, and neutral atom traps \cite{q17}.

But it is not easy to implement DTQW in circuit QED. Unlike other
systems, the superconducting qubits in circuit QED can not move. The
encoding of states of the coin and the walker in superconducting
qubits is difficult. There are only a few DTQW schemes in circuit
QED. In \cite{q18,q14}, the phase space of superposition states in a
cavity is used to encode the walker's position and a coupled qubit
is used as the coin. Due to the adopted non-orthogonal states of the
cavity and the limitation of the phase space, the generality and the
scalability of this scheme are inevitable problems. In \cite{q19}, a
1D DTQW scheme is proposed, in which a pair of superconducting
qubits are used as a node and nearest-neighbor qubits are coupled
via tunable couplers. The walker moves in the 1D line of nodes and
the coin is encoded by the position of the occupied qubit in each
node. In this way, an $N$-step DTQW needs at least $4N+2$ qubits and
$4N+1$ couplers.

In recent years, quantum information processing with qudits (
$d$-level systems), including qutrits (i.e., $d=3$), has been
attracting increasing interest, since qudits (with $d > 2$) can be
used to encode more information. For example, quantum information
processing and tomography of nanoscale semiconductor devices were
studied \cite{q20,q21}. In \cite{q22,q23}, schemes for quantum state
transfer of a qutrit in circuit QED were proposed.

In this paper, we propose an efficient and simple protocol for
implementation of DTQW in circuit QED, in which only $N+1$ qutrits
(with 3 energy levels) and $N$ cavities (i.e., the couplers) are
needed for an $N$-step DTQW. Since only resonant processes are
adopted, the operation of each DTQW step is very quick. With this
protocol, arbitrary initial states of the coin can be prepared and
arbitrary operation of the coin can be implemented easily. To
estimate the implementation of this protocol and the effects of
parameters, we numerically simulate the DTQW in a superconducting
system with the number of step up to $20$. It indicates that this
protocol is feasible with the present circuit QED technology and can
be used to implement a large-step DTQW. Because of the scalability
and rapid improvement of circuit QED technology, this protocol can
help to study properties and applications of large-step DTQW in
experiments, which is important for the development of quantum
information science.

\begin{figure}[t]
\begin{center}
\includegraphics[viewport=73 270 888 375, width=13.5cm]{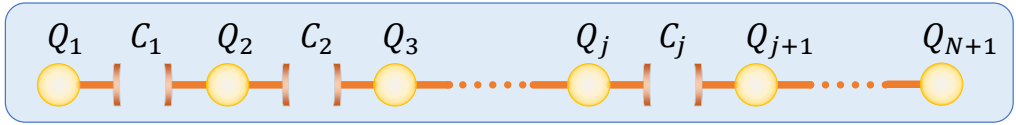}\vspace*{%
-0.00in}
\end{center}
\caption{(color online) Setup for implementation of DTQW in circuit
QED, which consists of $N+1$ qutrits and $N$ cavities.} \label{fig1}
\end{figure}

\section{A 1D DTQW protocol}

As shown in Fig.~\ref{fig1}, the setup consists of $N+1$ qutrits
(with energy levels $|g\rangle$, $|e\rangle$ and $|f\rangle$) and
$N$ cavities. All qutrits have the same energy levels and the
frequency of cavities ($\omega_c$) is equal to the
$|g\rangle\leftrightarrow|e\rangle$ transition frequency of the
qutrits ($\omega_{eg}$). The walker's position is represented by the
position of the qutrit in a non-ground state (i.e., superposition
state of $|e\rangle$ and $|f\rangle$), and the coin states
$|0\rangle_c$ and $|1\rangle_c$ are represented by the states
$|f\rangle$ and $|e\rangle$, respectively. In this case, arbitrary
initial state of the coin can be prepared easily by applying
corresponding pulses to the qutrits. Suppose all cavities are
initially in the ground state $|0\rangle$ and decoupled with
qutrits,
the steps for the implementation of the DTQW are as follows.


Step \textsf{I}: Tossing the coin by applying a pulse (with the Rabi
frequency $\Omega$) to each qutrit. The frequency, duration, initial
phase of the pulses are $\omega_{fe},t_{\sf{I}},\phi$, respectively.
In the interaction picture, the Hamiltonian for qutrits interacting
with the pulses is
$$H_{I,1}=\sum_j\Omega(e^{i\phi}|e\rangle_j\langle f|+e^{-i\phi}|f\rangle_j\langle e|).$$
This Hamiltonian makes the following transformations for states of
qutrit $j$ ($j=1,2,3,...,N+1$):
\begin{align}
&|g\rangle_{j}\rightarrow|g\rangle_{j},\nonumber\\
&|e\rangle_{j}\rightarrow\cos(\Omega t_{\sf{I}})|e\rangle_{j}-ie^{-i\phi}\sin(\Omega t_{\sf{I}})|f\rangle_{j},\nonumber\\
&|f\rangle_{j}\rightarrow-ie^{i\phi}\sin(\Omega
t_{\sf{I}})|e\rangle_{j}+\cos(\Omega
t_{\sf{I}})|f\rangle_{j}.\label{e1}
\end{align}
It shows that arbitrary unitary operator of the coin can be achieved
by applying suitable pulses. If we set $t_{\sf{I}}=\theta/\Omega$
and $\phi=-\pi/2$, this operation of the coin is just that of
$\sigma_z\cdot C$, where $C$ is the coin operator in
Eq.~(\ref{coin}) and
$\sigma_z=|0\rangle_c\langle0|-|1\rangle_c\langle1|$. The shift
operation of the walker with respect to the coin state will be
accomplished by the following two steps.

\begin{figure}[tb]
\begin{center}
\includegraphics[width=13.5cm,clip]{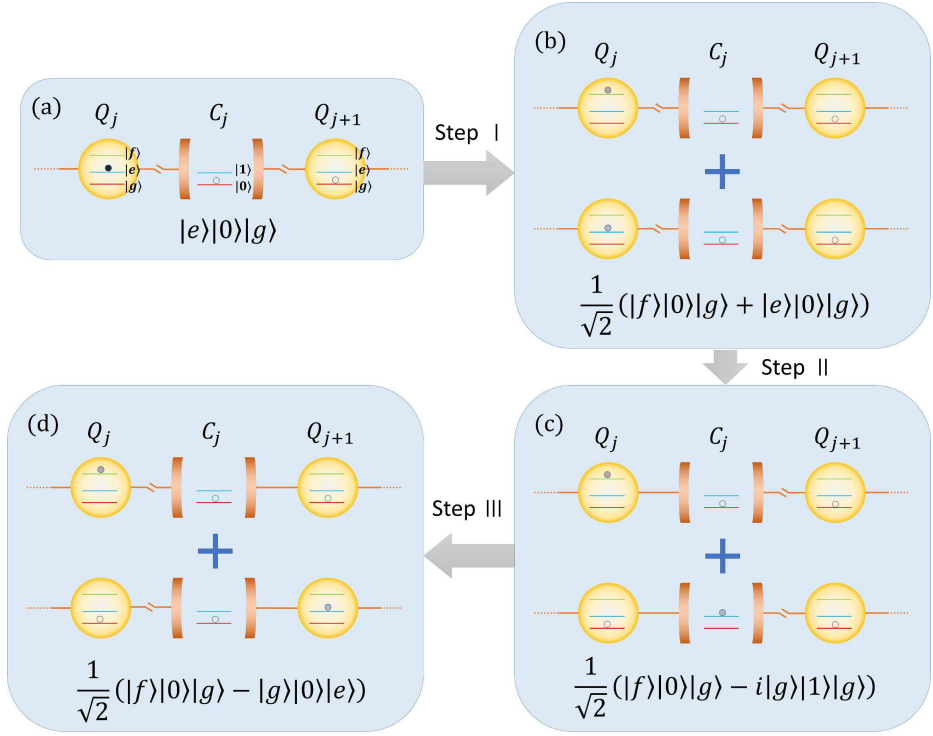} \vspace*{%
-0.6in}
\end{center}
\caption{(color online) Illustration of implementing steps of DTQW
and state evolutions by assuming that energy level $|e\rangle$ of
the qutrit $j$ is initially occupied (i.e., initially the walker is
at the position $j$ and the coin state is $|1\rangle$). The small
circles indicate the occupied energy levels. Hollow circles
represent the occupation of ground states and the shade of solid
circles represent the probability of the occupation of non-ground
states. $\theta=\pi/4$ has been assumed for the coin operator $C$.}
\label{fig2}
\end{figure}

Step \textsf{II}: Coupling cavity $j$ with the qutrit $j$ (assuming
the coupling strength $g_j=g$ with $j=1,2,3,...,N$ ). The
Hamiltonian in the interaction picture is
$$H_{I,2}=\sum_jg(a_j|e\rangle_j\langle g| + a^+_j|g\rangle_j\langle e|).$$
After the evolution time $t_{\sf{II}}=\pi/2g$, the transformations
for the states of the qutrit $j$ and the cavity $j$ are:
\begin{align}
&|g\rangle_{j}|0\rangle_{j}\rightarrow|g\rangle_{j}|0\rangle_{j},\nonumber\\
&|e\rangle_{j}|0\rangle_{j}\rightarrow-i|g\rangle_{j}|1\rangle_{j},\nonumber\\
&|f\rangle_{j}|0\rangle_{j}\rightarrow|f\rangle_{j}|0\rangle_{j}.\label{e2}
\end{align}
Step \textsf{III}: Decoupling cavity $j$ from the qutrit $j$ and
coupling cavity $j$ with the qutrit $j+1$ (assuming the coupling
strength $g'_j=\mu$ with $j=1,2,3,...,N$). Now the Hamiltonian in
the interaction picture becomes
$$H_{I,3}=\sum_j\mu(a_j|e\rangle_{j+1}\langle g| + a^+_j|g\rangle_{j+1}\langle e|).$$
After the evolution time $t_{\sf{III}}=\pi/2\mu$, the
transformations for the states of the cavity $j$ and the qutrit
$j+1$ are:
\begin{align}
&|0\rangle_{j}|g\rangle_{j+1}\rightarrow|0\rangle_{j}|g\rangle_{j+1},\nonumber\\
&|1\rangle_{j}|g\rangle_{j+1}\rightarrow-i|0\rangle_{j}|e\rangle_{j+1}.\label{e3}
\end{align}

If the walker is initially in position $j$ (i.e., the qutrit $j$ is
excited), after the step \textsf{II} and step \textsf{III}, it will
move into position $j+1$ (i.e., the qutrit $j+1$) with the coin
state $|1\rangle_c$ ($|e\rangle_{j+1}$) or it will stay in position
$j$ (i.e., the qutrit $j$) with the coin state $|0\rangle_c$
($|f\rangle_j$). This shift operation of the walker achieved in step
\textsf{II,III} can be expressed as $W'\cdot\sigma_z$, where
$W'=\sum_j( |j\rangle\langle
j|\cdot|0\rangle_c\langle0|+|j+1\rangle\langle
j|\cdot|1\rangle_c\langle1| )$. Note that $W'$ is equivalent to the
operator $W$ in Eq.(\ref{walker}).

Now the standard DTQW operation $U'=W'\cdot C$ is achieved with the
operational time $t=t_{\sf{I}}+t_{\sf{II}}+t_{\sf{III}}$. In
Fig.~\ref{fig2}, the evolutions of the states of the qutrits $j$,
$j+1$ and the cavity $j$ with steps \textsf{I},\textsf{II} and
\textsf{III} are demonstrated by assuming an initial state of
$|j\rangle|1\rangle_c$ (i.e., $|e\rangle_j$). Repeating the step
\textsf{I,II,III} by $N$ times, an $N$-step 1D DTQW is realized.


\section{POSSIBLE EXPERIMENTAL IMPLEMENTATION}

\begin{figure}[b]
\begin{center}
\includegraphics[bb=0 0 535 350, width=12cm]{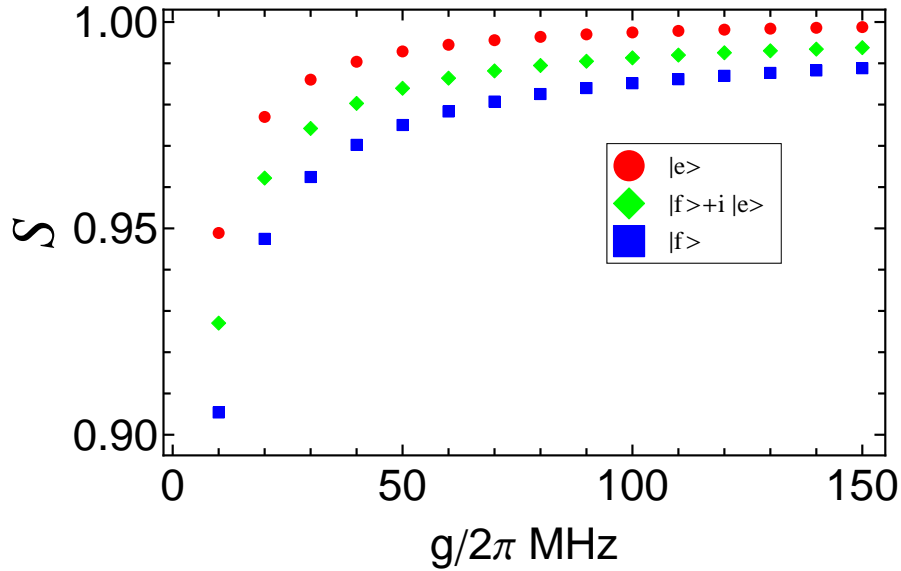}\vspace*{%
-0.4in}
\end{center}
\caption{(color online) Similarity versus $g/2\pi $ are plotted
respectively for initial states of the coin $|0\rangle_c,
|1\rangle_c, (|0\rangle_c+i|1\rangle_c)/\sqrt{2}$, with $N=10$ and
$\Omega/2\pi=100$ MHz.} \label{fig3}
\end{figure}

In this section, we discuss the feasibility for the implementation
of this DTQW protocol with DTQW steps up to $20$ by numerical
simulations. In all simulations, we set $\theta=\pi/4$ for the coin
operator and assume that the walker starts from the qutrit $1$. By
considering dissipation and dephasing, the evolving of the system is
determined by the master equation
\begin{eqnarray}
\frac{d\rho }{dt} &=&-i\left[H_{I,k},\rho \right]
+\sum_{j}\kappa _{a_{j}}\mathcal{L}\left[ a_{j}\right]  \notag \\
&&+\sum_{j}\gamma _{ef,j}\mathcal{L}\left[ \sigma _{ef,j}^{-}\right]
+\gamma _{gf,j}\mathcal{L}\left[ \sigma _{gf,j}^{-}\right] +\gamma _{ge,j}%
\mathcal{L}\left[ \sigma _{ge,j}^{-}\right]  \notag \\
&&+\sum_j\gamma _{e\varphi ,j}\mathcal{L}\left[\sigma
_{ee,j}\right]+\gamma _{f\varphi ,j}\mathcal{L}\left[\sigma
_{ff,j}\right], \label{me}
\end{eqnarray}%
where $\mathcal{L}\left[ \Lambda \right] =\Lambda \rho \Lambda
^{+}-\Lambda ^{+}\Lambda \rho /2-\rho \Lambda ^{+}\Lambda /2$ (with
$\Lambda =a_{j},\sigma _{ef,j}^{-},\sigma _{gf,j}^{-},\sigma
_{ge,j}^{-},\sigma _{ee,j},\sigma _{ff,j}$),\ $\sigma
_{ef,j}^{-}=\left\vert e\right\rangle _{j}\left\langle f\right\vert
,$ $\sigma
_{gf,j}^{-}=\left\vert g\right\rangle _{j}\left\langle f\right\vert ,$ $%
\sigma _{ge,j}^{-}=\left\vert g\right\rangle _{j}\left\langle
e\right\vert ,$ $\sigma _{ee,j}=\left\vert e\right\rangle
_{j}\left\langle e\right\vert $, and $\sigma _{ff,j}=\left\vert
f\right\rangle _{j}\left\langle f\right\vert
; $ $\kappa _{a_{j}}$ is the decay rate of cavity $j$;\ $%
\gamma _{ef,j}$ ($\gamma _{gf,j}$) is the energy relaxation rate for
the
level $\left\vert f\right\rangle $\ associated with the decay path $%
\left\vert f\right\rangle \rightarrow \left\vert e\right\rangle $ ($%
\left\vert f\right\rangle \rightarrow \left\vert g\right\rangle $)
of qutrit
$j$; $\gamma _{ge,j}$\ is the energy relaxation rate of the level $%
\left\vert e\right\rangle ;$ and $\gamma _{f\varphi ,j}$ ($\gamma
_{e\varphi,j}$) is the dephasing rate of the level $\left\vert f\right\rangle $ ($%
\left\vert e\right\rangle $) of qutrit $j$.

\begin{figure}[tb]
\begin{center}
\includegraphics[bb=0 0 535 350, width=12cm]{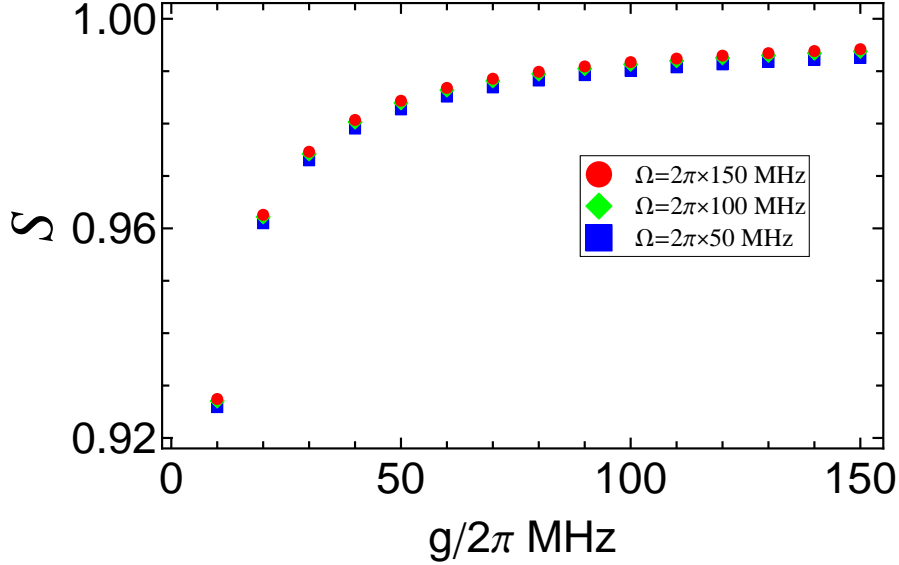}\vspace*{%
-0.4in}
\end{center}
\caption{(color online) Similarity versus $g/2\pi $ are plotted
respectively for $\Omega/2\pi=50, 100, 150$ MHz, with $N=10$ and the
initial state of the coin $(|0\rangle_c+i|1\rangle_c)/\sqrt{2}$.}
\label{fig4}
\end{figure}

In actual experiments, the probability distribution $P(j)$ of the
walker is always easy to measure. So we calculate the similarity
$S=\left(\sum_j\sqrt{P_{me}(j)P_{id}(j)}\right)^2$ to compare the
$P_{me}(j)$ from the master equation (\ref{me}) with the ideal
$P_{id}(j)$ of the standard DTQW. In numerical simulations, flux
qutrits are adopted and the decoherence parameters used are: (i)
$\gamma _{e\varphi ,j}^{-1}=5$ $\mu $s$,$
$\gamma _{f\varphi ,j}^{-1}=5$ $\mu $s; (ii) $\gamma _{ge,j}^{-1}=10$ $\mu $%
s, $\gamma _{ef,j}^{-1}=10$ $\mu $s, $\gamma _{gf,j}^{-1}=10$ $\mu $s \cite%
{s16,s17,s18,s19,s20}, and (iii) $\kappa _{a_{j}}^{-1}=10$ $\mu $s
\cite{s22,s23}. We denote this set of the decoherence parameters as
$T$ and the above choice of $T$ as $T_0$. For simplicity, we will
set the coupling strengths $g_j=g'_j=g$. We will study the effects
of the coupling strength $g$, Rabi frequency $\Omega$, initial state
of the coin $|\phi_{c0}\rangle$, the number of DTQW steps $N$ and
the decoherence time set $T$ on the similarity $S$.

\begin{figure}[tb]
\begin{center}
\includegraphics[bb=0 0 535 350, width=12cm]{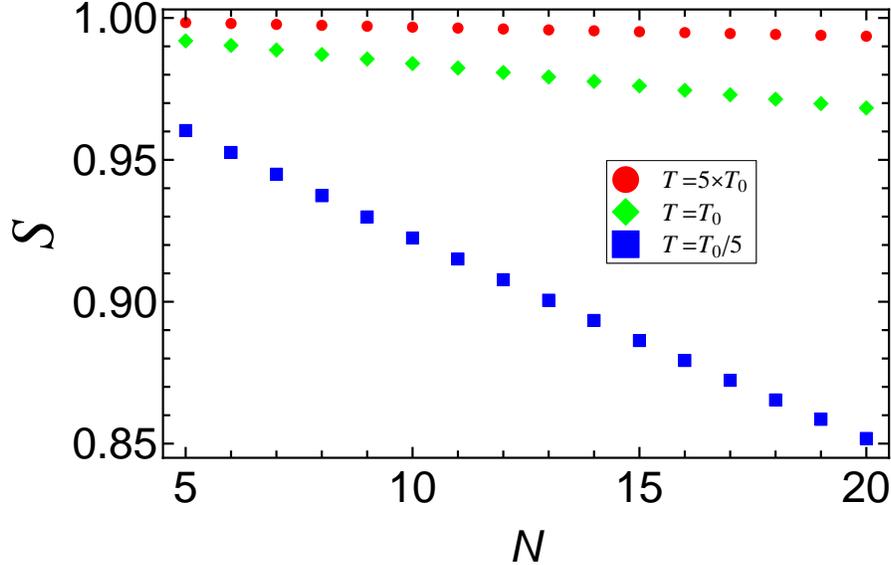}\vspace*{%
-0.4in}
\end{center}
\caption{(color online) Similarity versus number of step $N$ are
plotted respectively for $T=5T_0, T_0, T_0/5$, with $g/2\pi=50$ MHz,
$\Omega/2\pi=100$ MHz and the initial state of the coin
$(|0\rangle_c+i|1\rangle_c)/\sqrt{2}$.} \label{fig5}
\end{figure}

In Fig.~3, the similarity versus $g/2\pi $ is plotted respectively
for $|\phi_{c0}\rangle=|0\rangle_c$, $|1\rangle_c$ and
$(|0\rangle_c+i|1\rangle_c)/\sqrt{2}$ (i.e., $|f\rangle$,
$|e\rangle$ and $(|f\rangle+i|e\rangle)/\sqrt{2}$), with $N=10$ and
$\Omega/2\pi=100$ MHz \cite{s24}. It is shown that different initial
states of the coin lead to different similarities. Since each
initial state of the coin is a superposition of $|0\rangle_c$ and
$|1\rangle_c$ (i.e., $|f\rangle$ and $|e\rangle$), the lowest
similarity for all initial states should be that with the initial
state $|0\rangle_c$ (i.e., $|f\rangle$). As $g/2\pi=50$ MHz, it
shows that the similarity $S$ is larger than $0.975$ for an
arbitrary initial state of the coin in the 10-step DTQW. Moreover,
the similarity increases with the increasing of $g$ as expected. It
is shown that high similarity of this protocol can be achieved with
present technology of circuit QED.

In Fig.~4, similarity versus $g/2\pi $ are plotted respectively for
$\Omega/2\pi=50, 100, 150$ MHz, with $N=10$ and
$|\phi_{c0}\rangle=(|0\rangle_c+i|1\rangle_c)/\sqrt{2}$. It shows
that the similarity is not sensitive to the Rabi frequency $\Omega$.

In Fig.~5, similarity versus number of step $N$ are plotted
respectively for $T=5T_0, T_0, T_0/5$, with $g/2\pi=50$ MHz,
$\Omega/2\pi=100$ MHz and
$|\phi_{c0}\rangle=(|0\rangle_c+i|1\rangle_c)/\sqrt{2}$. It
indicates that similarity decreases with the increasing of $N$ and
with the decreasing of $T$ as expected. For larger $T$, similarity
decreases more slowly with the increasing of the step $N$. For
larger $N$, similarity is more sensitive to the variance of $T$. For
$N=20$, we obtain similarity $S\sim0.993, 0.968, 0.852$ with
$T=5T_0, T_0, T_0/5$, respectively. If the coherent times of the
devices are not very small, high similarity for a large step DTQW
can still be achieved with this DTQW protocol in circuit QED.

In Fig.~6, we show the probability distribution of the DTQW from the
numerical simulation and that of the ideal DTQW, with $N=20$,
$T=T_0$, $g/2\pi=50$ MHz, $\Omega/2\pi=100$ MHz and the initial
state of the coin $(|0\rangle_c+i|1\rangle_c)/\sqrt{2}$. In this
case, the similarity $S$ is $\sim0.968$ and the numerical simulation
indicates that probability distribution of the DTQW implemented in
circuit QED with number of steps up to 20 still meets well with that
of ideal DTQW.

\begin{figure}[tb]
\begin{center}
\includegraphics[bb=0 0 750 485, width=12cm]{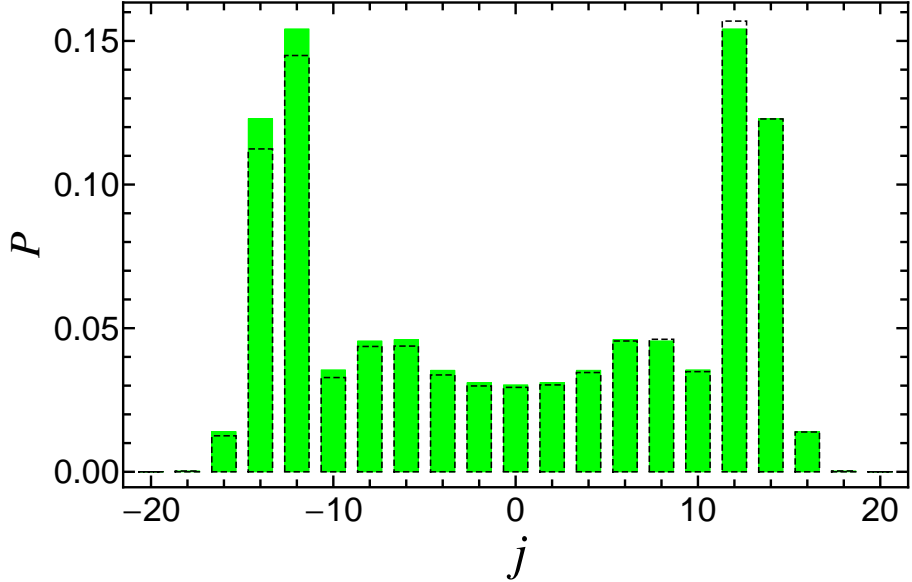}\vspace*{%
-0.4in}
\end{center}
\caption{(color online) Probability distribution of the DTQW from
the numerical simulation (dashed bars) compared with that of the
ideal DTQW (green bars), with $N=20$, $T=T_0$, $g/2\pi=50$ MHz,
$\Omega/2\pi=100$ MHz and the initial state of the coin
$(|0\rangle_c+i|1\rangle_c)/\sqrt{2}$.} \label{fig6}
\end{figure}

\section{CONCLUSIONS}

We have presented a protocol for implementing standard DTQW in
circuit QED. The protocol is simple and efficient, only $N+1$
qutrits and $N$ assistant cavities are needed for an $N$-step DTQW
and the operational time for each step is rather short due to the
adoption of resonant processes. With this protocol, arbitrary
initial states of the coin can be prepared and arbitrary operation
of the coin can be implemented easily, which is necessary for the
general researches and applications of DTQW. The numerical
simulations prove that high-similarity DTQW with $N\leqslant 20$ is
feasible with present-day circuit QED technique. This DTQW protocol
is quite general and can be extended to implement multi-dimensional
DTQW in circuit QED, which is important in the quantum computation
and quantum simulation.

\section*{ACKNOWLEDGMENTS}
This work was supported in part by the NKRDP of China (Grant No.
2016YFA0301802), the National Natural Science Foundation of China
under Grant Nos. [11504075, 11074062, 11374083].

\end{document}